# Observation of dynamic charge stripes in $Tm_{0.19}Yb_{0.81}B_{12}$ at the metal-insulator transition.

N. E. Sluchanko[a,b*], A. N. Azarevich[a,b], A. V. Bogach[a], N. B. Bolotina[c], V. V. Glushkov[a,b], S. V. Demishev[a,b], A. P. Dudka[a,c], O. N. Khrykina[a,c], V. B. Filipov[d], N. Yu. Shitsevalova[d], G. A. Komandin[a], A.V. Muratov[e], Yu. A. Aleshchenko[e], E. S. Zhukova[b], B. P. Gorshunov[a,b]

[a] *Prokhorov General Physics Institute, Russian Academy of Sciences, Vavilova 38, Moscow, 119991 Russia*

[b] *Moscow Institute of Physics and Technology, 9, Institutskii per., Dolgoprudnyi, Moscow region, 141700 Russia*

[c] *Shubnikov Institute of Crystallography of Federal Scientific Research Centre Crystallography and Photonics of Russian Academy of Sciences, 59 Leninskii av., Moscow, 119333 Russia*

[d] *Frantsevich Institute for Problems of Materials Science, National Academy of Sciences of Ukraine, 3, Krzhyzhanovsky str., Kiev, 03680 Ukraine*

[e] *Lebedev Physical Institute, Russian Academy of Sciences, 53 Leninskii av., Moscow, 119991 Russia*

\* -E-mail: nes@lt.gpi.ru

**Abstract.** Higher accuracy low temperature charge transport measurements in combination with precise X-ray diffraction experiment have allowed detecting the symmetry lowering in the single domain $Tm_{0.19}Yb_{0.81}B_{12}$ crystals of the family of dodecaborides with metal-insulator transition. Basing on the fine structure analysis we discover formation of dynamic charge stripes within the semiconducting matrix of $Tm_{0.19}Yb_{0.81}B_{12}$. The charge dynamics in these metallic nano-size conducting channels is characterized by broad-band optical spectroscopy that allowed estimating the frequency (~$2.4 \cdot 10^{11}$ Hz) of quantum motion of the charge carriers. It is suggested that caused by cooperative Jahn-Teller effect in the boron sub-lattice, the large amplitude rattling modes of the Tm and Yb ions are responsible for 'modulation' of the conduction band along [110] direction through the variation of *5d-2p* hybridization of electron states.

**Introduction.** Dynamic or fluctuating charge and spin stripes attract extraordinary scientific interest due to their possible role in high temperature superconductivity mechanism [1-14]. While stripe states were previously thought to be a special feature of the $La_{2-x}Ba_xCuO_4$ family of perovskites [15], singularities of phases with lowered lattice symmetry have been detected in a number of other cuprates including $YBa_2Cu_3O_{7-\delta}$ [16-17], $SmBa_2Cu_3O_x$ [18], $Bi_2Sr_2CaCu_2O_{8+\delta}$ [19] and $Ca_{2-x}Na_xCuO_2Cl_2$ [20]. This kind of electronic instability is also suggested to be extremely important for understanding of physics of colossal magnetoresistive manganites [21-22], nickelates [23-25], iron-based superconductors [26-27], heavy fermion hexaborides [28], rare earth dodecaborides [29], etc.

During last twenty years a special attention was paid to search for an answer on two main questions: (1) Is there any universal scenario of the dynamic charge and spin stripe formation in the strongly correlated electron systems? and (2) How can the dynamic stripes be detected experimentally? The origin of conducting stripes is discussed usually from a strong-coupling perspective and various approaches were developed to describe (*i*) frustrated phase separation (or micro phase separation), (*ii*) spin-charge "topological" properties, and (*iii*) valence-bond solid formation [14]. Although the existence of both charge stripes and nematic order in strongly correlated electron systems is now a well-established experimental fact, the origin of these inhomogeneities remains controversial. Taking into account that, from the strong-coupling perspective, stripes are a real-space pattern of micro phase separation [8], the nanoscale visualization of the filamentary structure could be realized with the help of scanning tunneling microscopy and spectroscopy (STM and STS). However, STS and STM are essentially static and surface sensitive techniques and consequently dynamic stripes can only be detected if pinned by impurities. Among indirect methods, anisotropic charge transport measurements on single domain crystals are considered as most effective technique enable to detect the symmetry lowering below the transition to fluctuating stripe phase [14]. Among other experiments that recently have been successfully used for observation of dynamic charge stripes, both inelastic neutron scattering [30] and ultrafast and equilibrium mid-infrared spectroscopy [31] should be mentioned. In these studies the lattice fluctuations associated with the dynamic charge stripes were detected in the model compounds - doped nickelates $La_{2-x}Sr_xNiO_4$ (see also [32]).

Very recently, fluctuating charge stripes have been detected in non-magnetic dodecaboride $LuB_{12}$ [29,33] and their origin was explained in terms of the cooperative dynamic Jahn-Teller effect in the boron $B_{12}$cubooctahedra. It was suggested [29,33], that this kind of boron lattice instability induces large amplitude vibrations of rare earth ions (rattling modes) resulting in (*i*) dramatic changes of $5d - 2p$ hybridization of electron states along the unique crystallographic axis and (*ii*) formation of highly conductive channels—dynamic charge stripes with a strong charge carrier scattering on the filamentary structure in the metallic matrix of $RB_{12}$. Taking into account that these *metallic* channels in the rare earth dodecaborides could be better visualized when formed within lower-conducting, dielectric matrix, it looks promising to investigate the stripe signatures in the crystals with composition close to $YbB_{12}$ that is a narrow-gap semiconductor, namely, the solid solutions of the $Tm_{1-x}Yb_xB_{12}$ family [34, 35]. That is why we have chosen for such studies crystals of $Tm_{1-x}Yb_xB_{12}$ with *x*=0.81 and with isotopically pure boron ($^{11}B$) to avoid any $^{10}B$-$^{11}B$ substitutional disorder. Below we present the results of low temperature resistivity, magnetoresistance and transverse even effect measurements in combination with both precise X-ray diffraction and broad-band room-temperature optical studies. It is shown that dynamic charge stripes in $Tm_{1-x}Yb_xB_{12}$ may be reliably detected in the difference Fourier maps resulting from the accurate analysis of the crystal structure fine details. The lower symmetry electron-density distribution is confirmed by the maximal entropy method (MEM) [36]. At low temperatures the filamentary structure is also detected in the measurements of anisotropy of both the magnetoresistance and transverse even effect [35]. Moreover, about an order of magnitude different DC- and far-infrared AC-conductivities observed at room temperature are also indicative of the dynamic stripe formation in the **$Tm_{0.19}Yb_{0.81}B_{12}$** dodecaboride matrix.

**Experimental details.** The details of growth of $Tm_{1-x}Yb_xB_{12}$ single crystals, preparation of samples and the technique of magnetotransport measurements on the installation with a sample rotation in a magnetic field were described in [35, 37]. In the present study we have used the same X-rays diffractometers and special software to extract fine details of the crystal structure as it was done recently with $LuB_{12}$ [29,33]. Broad-band reflectivity measurements of $Tm_{0.19}Yb_{0.81}B_{12}$ were carried out on the high quality $Tm_{0.19}Yb_{0.81}B_{12}$ single crystal ≈6 mm in diameter using a set of apparatus, including Vertex 80V Fourier-transform spectrometer, J.A. Woollam V-VASE ellipsometer, cw backward-wave oscillator [38] and pulsed TeraView THz spectrometers (see [39] for more details) and the data from [40] were used to extend the reflection coefficient spectra up to ≈400000 cm$^{-1}$.

**Results and discussion.** *Magnetotransport anisotropy*. Fig. 1a shows the temperature dependences of the resistivity $\rho(T)$ and the Hall coefficient $R_H(T, H_0)$ of the $Tm_{0.19}Yb_{0.81}B_{12}$ samples. The Arrhenius plot of the data (see [35] for details) reveals two activation regimes on the $R_H(T, H_0)$ curve at T> 10 K, corresponding to the energy gap $E_g$≈ 200 K≈ 17.8 meV (I) and the binding energy of the manybody states (resonance at $E_F$) in the gap $E_p$≈ 75 K (II), respectively. At temperatures below 10 K a transition to a coherent regime of charge transport is observed in the low field region that is suppressed by an external magnetic field of 80 kOe (Fig. 1a, intervals I-III, respectively). The low values of the Hall mobility (3-24 cm$^2$/(V s), see the inset in Fig. 1a) indicate a very strong scattering of charge carriers on charge and spin fluctuations in $Tm_{0.19}Yb_{0.81}B_{12}$.

Measurements of charge transport anisotropy at a temperature of 2 K in a coherent regime are performed in a scheme corresponding to the transverse magnetoresistance and Hall resistance which are recorded when the crystal rotates around the current axis (**I** //[110] ⊥ **H**, see fig.1b where schematic arrangement of these angular measurements is shown). It is worth noting that in addition to the ordinary Hall effect signal $\rho_H$(T, $H_0$) in $Tm_{0.19}Yb_{0.81}B_{12}$ a transverse even effect $\rho_{TE}(\varphi, H_0)$ is also recorded from the Hall probes, and this second harmonic term $\rho_{TE}(\varphi, H_0)= \rho_{H2}\cos2\varphi$ is directly related to the formation of the filamentary structure of the conducting channels in the semiconducting matrix of the dodecaboride [35]. The procedure of separating the isotropic odd Hall component $\rho_H(\varphi, H_0)$ and the anisotropic transverse even contribution $\rho_{TE}$ is described in detail in [35]. The families of both resistivity curves $\rho_{anis}/\rho_0(\varphi,H_0)$ and $\rho_{TE}(\varphi, H_0)$ separated within the analysis (see also [39], for details) are presented in polar coordinates in figs.1c and 1d, respectively (color shows the amplitude of the resistivity components). It is clearly seen that these two carrier scattering diagrams demonstrate uniaxial anisotropy and the selected axis is located predominantly along the **H**//[110] direction for both the resistivity (Fig.1c) and the transverse even effect (Fig.1d).

*Crystal anisotropy*. The compound $Tm_{0.19}Yb_{0.81}B_{12}$ crystallizes in a cubic lattice with the period $a$ = 7.4679(3). We refined the structure in the high-symmetry *Fm-3m* group using room temperature X-ray data with low residual factor $R$ = 0.49%. At the same time, accurately measured periods of the crystal lattice appeared to be slightly different: $a$ = 7.4686(5), $b$ = 7.4684(4), $c$ = 7.4668(4). Similar 'tetragonal' distortion a ≈ b > c was observed in $LuB_{12}$ [29]. Difference Fourier synthesis of electron density (ED) [39], which in fact needs no data on crystal symmetry, is used to reveal fine structural details. Difference Fourier maps (Fig. 2a) show a low-symmetry distribution of difference ED in the faces (100), (010), (001) of the unit cell. The asymmetry of ED distribution is confirmed by processing the structural data using the maximal entropy method (MEM) [36], which does not need atomic coordinates and chemical composition to restore ED from observed structure factors $F_{obs}$. MEM maps in the three faces are presented in Fig. 2b for comparison. MEM reconstructs 'normal' but not difference ED, so light boron atoms are as clearly seen as heavy R atoms. As is clear from Fig. 2, the traces of conducting channels are observed along the face diagonals [011], [10-1], [110], but the most pronounced charge stripes are detected along [10-1] in the (010) plane.

*Dynamic conductivity.* The broad-band reflectivity spectrum of $Tm_{0.19}Yb_{0.81}B_{12}$ single crystal as presented in fig.3a (see [39] for measurement details). With the Kramers-Kronig analysis, the spectrum of dynamical conductivity was calculated, as shown in Fig.3b. Rich set of peaks in the spectrum was modeled with the sum of Lorentzian expressions $\sigma^*(\nu) = \frac{0.5\Delta\varepsilon\nu_0^2\nu}{\nu\gamma+i(\nu_0^2-\nu^2)}$ where $\Delta\varepsilon$ is the dielectric contribution, $\nu_0$ is resonance frequency and $\gamma$ is the damping constant. The details of the conductivity spectrum analysis and parameters of corresponding excitations will be published elsewhere (see also Table 1 in [39]). Among the most important results, two issues should be mentioned here: (*i*) the value of conductivity at terahertz frequencies, $\sigma$(30-40 cm$^{-1}$) ≈ 1500 $\Omega^{-1}$cm$^{-1}$, is by about an order of magnitude below the measured DC conductivity $\sigma_{DC}$≈13000 $\Omega^{-1}$cm$^{-1}$ and (*ii*) modeling the mismatch with the Drude conductivity term (dashed line in Fig.3b) $\sigma(\nu) = \sigma_{DC}(1 - i\nu/\gamma)^{-1}$ provides the scattering rate of carriers $\gamma$≈8 cm$^{-1}$ which coincides very well with the damping of two quasilocal vibrations of the heavy R-ions (R-Tm and Yb) located at 107 cm$^{-1}$ (~154 K) and 132 cm$^{-1}$ (~190 K) (see Table 1 in [39]). For comparison, similar values of Einstein temperature $\Theta_E$=160-206 K have been detected previously in EXAFS [41], heat capacity [42] and inelastic neutron scattering [43] studies of the dodecaborides. Taking into account that these rattling modes induce variation of *5d-2p* hybridization in $RB_{12}$ [29], we suggest that this kind of "modulation" of the conduction band is responsible for the dynamic charge stripes formation in the dodecaboride matrix. Besides, the frequency ~2.4·10$^{11}$ Hz (8 cm$^{-1}$) may be deduced from the conductivity spectra as the characteristic of the quantum motion of charges in the dynamic stripes. As a result, the DC conductivity in $Tm_{0.19}Yb_{0.81}B_{12}$ at room temperature should be attributed to the charge transport in dynamic stripes that are more conductive than surrounding semiconducting matrix. Consequently, the value of the DC conductivity is determined by the stripes that percolate through the crystal while relatively smaller AC conductivity is provided by the THz-FIR reflectivity of the 'whole' sample (conducting stripes + semiconducting matrix).

**Conclusion.** A set of experimental results is obtained that evidence formation of dynamic charge stripes within the semiconducting matrix of $Tm_{0.19}Yb_{0.81}B_{12}$. The uniaxial anisotropy of the magnetoresistance and transverse even effect is observed in the single domain face-centered cubic $Tm_{0.19}Yb_{0.81}B_{12}$ crystals. Precise X-rays diffraction experiment and structural data analysis combined with the reconstruction of difference Fourier map of residual electron density and applying the maximum entropy method to deduce normal ED allowed to visualize the dynamic charge stripes oriented predominantly along the [110] direction with a slight inclination to the [111] axis. The charge dynamics in these stripes is characterized by the broad-band optical reflectivity measurements, which provide estimate of the frequency of the quantum motion of the charge carriers in the nano-size filamentary cannels. It is suggested that caused by the cooperative Jahn-Teller effect in the boron sub-lattice, the large-amplitude quasi-local vibrations of the Tm and Yb ions, rattling modes (Einstein oscillators at 107 $cm^{-1}$ and 132 $cm^{-1}$) are responsible for variation of *5d-2p* hybridization of electron states and hence for 'modulation' of the conduction band along the [110] direction, resulting in emergence of the dynamic charge stripes and symmetry lowering in the dodecaborides. Taking into account that many strongly correlated electron systems including CMR manganites and HTSC are inhomogeneous at the nanoscale, the observation of dynamic charge stripes in the $RB_{12}$ test bed compounds are crucial to unveiling the subtle nanoscale phase separation tendencies that induce a variety of real-space patterns [44].

**Acknowledgements.** This work was supported by the Russian Science Foundation, project no. 17-12-01426, and it was performed with the use of the equipment of the Shared Facility Centers of Lebedev Physical Institute of RAS (program 5-100) and FSRC Crystallography and Photonics of RAS.

**Figure captions.**

**Fig.1.** (a) Temperature dependences of resistivity $\rho(T)$ and the Hall coefficient $R_H(T, H_0)$ of $Tm_{0.19}Yb_{0.81}B_{12}$ for $H_0$=15 kOe and 80 kOe. Symbols I-III mark the temperature intervals (see text). Inset shows the temperature dependence of Hall mobility $\mu_H(T)=R_H(T, H_0)/\rho(T)$. (b) Schematic arrangement of the resistivity experiment with the sample rotation. **I** - measuring current, **n** - normal vector to the sample surface, angle $\varphi \equiv n\wedge H$, $U_H$ and $U_r$ are voltages from the Hall and potential probes to the crystal. The families of both resistivity curves $\rho_{anis}/\rho_0(\varphi,H_0)$ (c) and $\rho_{TE}(\varphi, H_0)$ (d) are presented in polar coordinates (color on line).

**Fig. 2.** Difference Fourier (a) and MEM (b) maps are built in (100), (010), (001) faces of the unit cell. ED in the layer of any given thickness is automatically divided into several levels from zero to $\rho_{max}$, each of them is assigned to a definite color from dark-blue over green to red. The values of $\rho_{MEM}$ are cut at the level $g_{max} = 0.075\%$ of a maximum $g_{MEM}(R)$ to show fine ED gradations in the thin layer. Difference ED values are cut at $\pm 0.5$ e/Å$^3$.

**Fig.3.** (a) Room temperature spectra of reflection coefficient of $Tm_{0.19}Yb_{0.81}B_{12}$ single crystal. Dots at 20 and 38 cm$^{-1}$ are reflectivity data measured with the backward-wave oscillator spectrometer. Dotted line corresponds to Hagen-Rubens reflectivity $R = 1 - \sqrt{4\nu/\sigma_{DC}}$ calculated with the measured $\sigma_{DC}$=13000 $\Omega^{-1}$cm$^{-1}$. Solid line is the result of fitting the spectrum with a set of Lorentzians as described in the text. Corresponding excitations are shown separately in panel (b) that shows spectrum of conductivity (black solid line) of $Tm_{0.19}Yb_{0.81}B_{12}$ obtained by Kramers-Kronig analysis of the reflectivity. Red dots above 4000 cm$^{-1}$ correspond to direct measurement on ellipsometer. Dashed line models the mismatch between the THz-FIR and DC conductivity with the Drude expression $\sigma(\nu) = \sigma_{DC}(1 - {i\nu}/{\gamma})^{-1}$ as described in the text. Stars denote quasilocal vibrations of heavy R-ions at 107 cm$^{-1}$ and 132 cm$^{-1}$ with the damping $\gamma \approx 8$ cm$^{-1}$ [39].

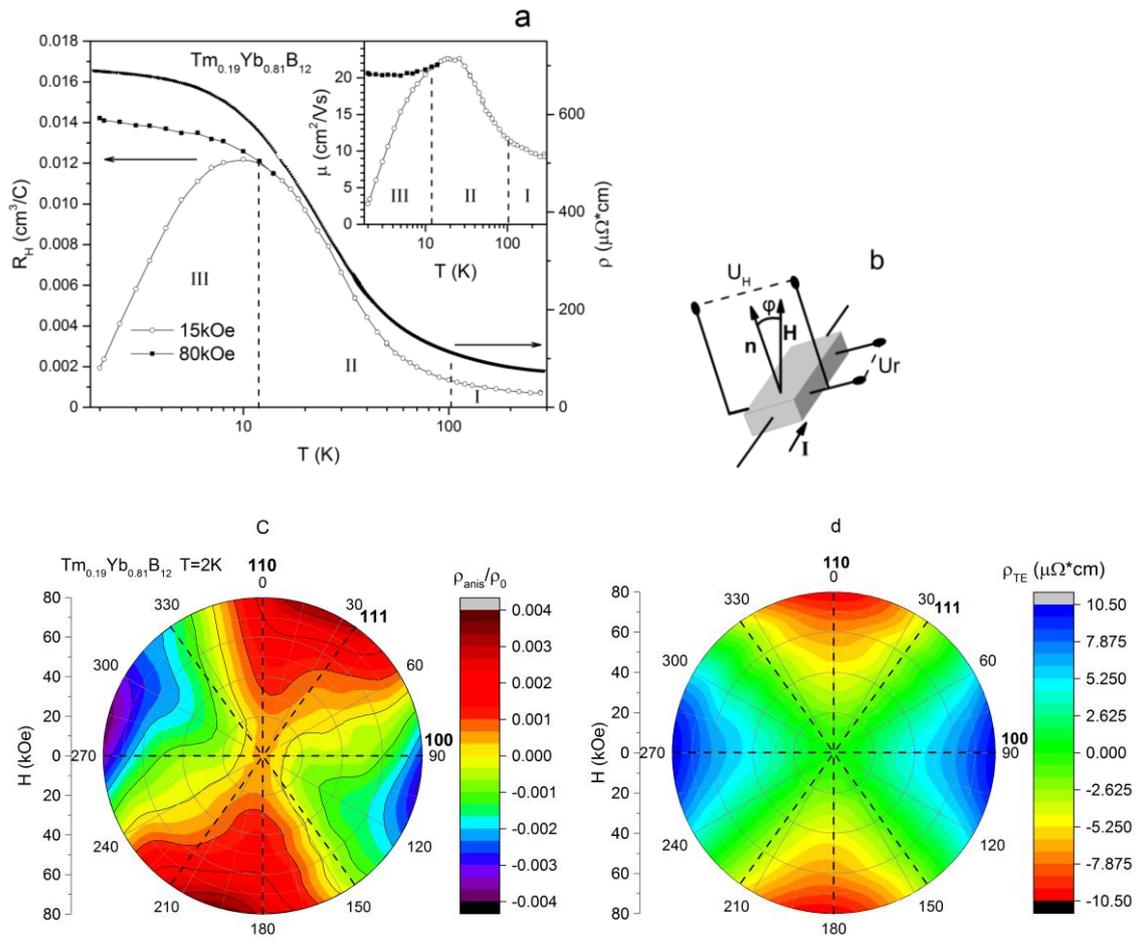

**Fig.1.**

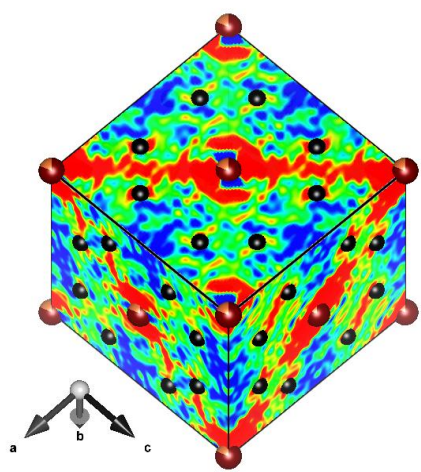
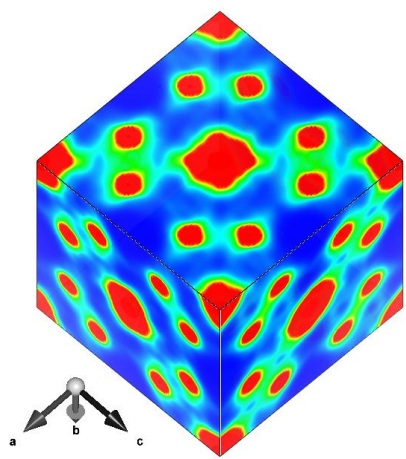

(a) (b)

**Fig.2.**

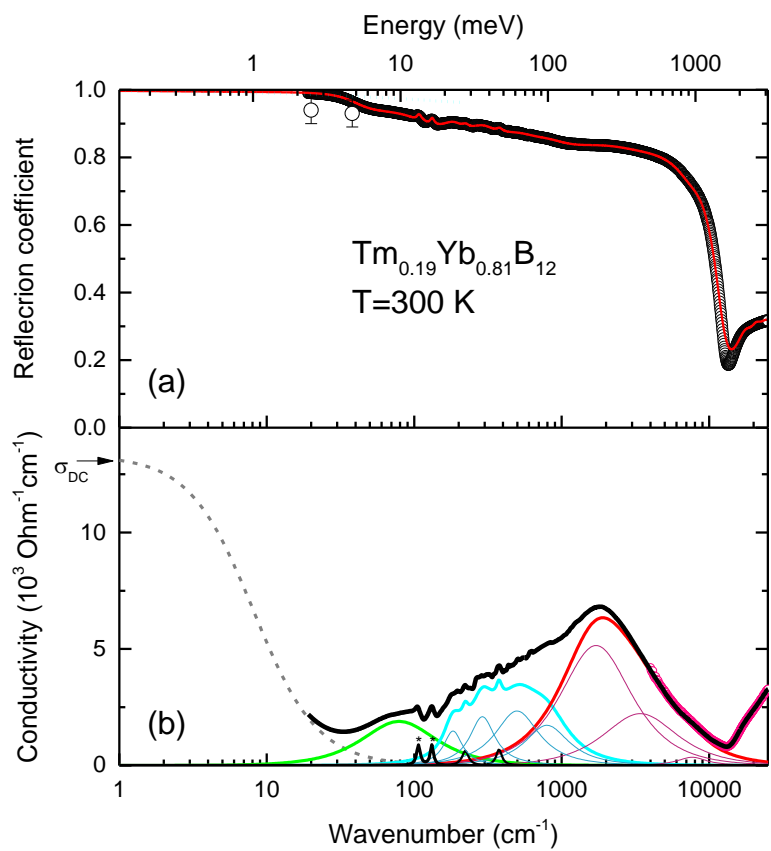

**Fig.3.**